\documentclass[twocolumn,twocolappendix]{aastex631}

\usepackage{amsmath}
\usepackage[utf8]{inputenc}
\shorttitle{Calibrating Eruptive Mass Loss in RSGs with Local Group Data}
\shortauthors{Cheng et al.}

\graphicspath{{./}{figures/}}

\begin{document}
	
	\title{Calibrating Eruptive Mass Loss in Red Supergiants with Local Group Data}

	\correspondingauthor{Shelley J. Cheng}
	\email{shelley.cheng@cfa.harvard.edu}
	
	\author[0000-0002-5462-0294]{Shelley J. Cheng}
	\affiliation{Center for Astrophysics $|$ Harvard \& Smithsonian, 60 Garden Street, Cambridge, MA 02138, USA}
	
	\author[0000-0002-1590-8551]{Charlie Conroy}
	\affiliation{Center for Astrophysics $|$ Harvard \& Smithsonian, 60 Garden Street, Cambridge, MA 02138, USA}
	
	\author[0000-0003-1012-3031]{Jared A. Goldberg}
    \affiliation{Department of Physics and Astronomy, Michigan State University, East Lansing, MI 48824, USA}
	\affiliation{Center for Computational Astrophysics, Flatiron Institute, 162 5th Avenue, New York, NY 10010, USA}
	
	
	

	\begin{abstract}

We calibrate a physically motivated, super-Eddington eruptive mass-loss prescription for red supergiants (RSGs) using Local Group stellar populations. Building on \texttt{MESA} models that add eruptive mass loss with a free scaling parameter $\xi$, we generate stellar evolution tracks and isochrones, and synthesize mock populations at metallicities of $Z/Z_\odot=0.2,\ 0.4$, and $1.0$. We compare model luminosity functions to observations of RSGs in the SMC, LMC, and M31, restricting to $3.5<\log T_{\rm eff}/K<3.75$ and $\log(L/L_\odot)>4.5$. By-eye fits to the observations yield values of $\xi_\mathrm{SMC}=0.0-0.05$, $\xi_\mathrm{LMC}=0.1$, and $\xi_\mathrm{M31}=0.35$, implying a positive, linear trend between the strength of eruptive mass-loss and metallicity. This calibrated eruptive mass loss prevents stars with initial masses $\gtrsim 20~M_\odot$ from evolving to become red supergiants, with implications for the mass spectrum of core-collapse progenitors, compact remnants, early supernova interaction signatures, and the spectral energy distributions of unresolved galaxies.

\end{abstract}
	
	
\section{Introduction} \label{sec:intro}

Red supergiants (RSGs) mark a pivotal late stage in the lives of massive stars ($\gtrsim 10\,M_\odot$) where their mass-loss rates can become large enough to alter subsequent evolution \citep{Meynet:2003,Ekstrom:12}. On galactic scales, populations of massive stars shape the structure, dynamics, and overall evolution of their hosts through radiation, winds, binary interaction, and eruptive mass loss, modifying gas densities and kinematics and driving feedback \citep[e.g.,][]{Tacconi:20,Heckman:1990,Ceverino:09,Geen:15,Steinwandel:2023}. On smaller scales, constraining intrinsic mass-loss histories is crucial for predicting the mass spectrum of compact remnants \citep{Belczynski:10,Dominik:12,Renzo:17,Banerjee:20,Moriya:2022,Fragos:23}, modeling the formation pathways of gravitational-wave sources \citep{Ziosi:14,Mapelli:16,Uchida:2019,Vink:21}, and interpreting the circumstellar environments that imprint early-time interaction signatures in supernova light curves and spectra \citep[e.g.,][]{Schlegel:1990,Pastorello:2007,Morozova2017,Bruch2023,JacobsonGalan2024}. More broadly, RSG mass loss affects the integrated optical/near-IR light of young populations, propagating into stellar-population synthesis models and the SEDs of unresolved galaxies \citep[e.g.,][]{Lancon:07,Conroy:10}.

Observational constraints on RSG mass loss have historically been challenging to obtain due to the complexity of measuring time-averaged mass-loss rates for evolved stars. Additionally, the emissivity of recombination emission lines scales with the square of the wind density, so even modest density inhomogeneities (clumping) can lead to overestimated $\dot{M}$ if not properly modeled.  Direct determinations from infrared excess and radio observations typically sample only the current mass-loss rate, which may not be representative of the time-averaged behavior \citep{Mauron:11,Ma:2025}. Alternative approaches have focused on statistical studies of RSG populations, using luminosity functions and number counts to infer average mass-loss properties \citep{Neugent:20,Massey:21}. These population-level studies offer the advantage of constraining time-averaged mass-loss rates by comparing observed stellar number distributions with theoretical predictions from stellar evolution models. A drawback is that population statistics average over star-to-star variability and can obscure episodic mass loss.

On the modeling side, stellar evolution calculations frequently implement massive star mass loss using empirical prescriptions \citep{Wood:83, deJager:88, Wood:92, Smith:2014, Meynet:2003, Nieuwenhuijzen:90, Vink:2001,vanLoon:05}. \citet{Mauron:11} and \citet{Renzo:17} provide comparisons among several commonly adopted prescriptions. 

Luminous blue variables can experience eruptive mass-loss with rates reaching $\dot{M}\sim 0.1$--$5~M_\odot~\mathrm{yr^{-1}}$ during rare giant eruption events such as those seen in $\eta$ Car \citep{Humphreys:1984, Smith:2017}; such phenomena are not captured by steady-wind prescriptions. More commonly observed S Doradus variability involves considerably lower mass-loss rates on the order of $\dot{M}\sim 10^{-4}~M_\odot~\mathrm{yr^{-1}}$   \citep[e.g.,][]{Vink:02,Smith:04}. There are a variety of models attempting to explain the origin of eruptive mass loss \citep[e.g.,][]{Owocki:2004,Smith:2006}, with recent work focusing on the role of radiation-driven winds operating near or above the Eddington limit \citep{Owocki:2004, Vink:21, Delbroek:2025, Jiang:15, Jiang:18, Cheng:24}.  In recognition of the potential importance of eruptive mass loss for stellar evolution, several groups have adopted simple schemes or 
empirically tuned prescriptions \citep[e.g.,][]{Ekstrom:12,Dessart:13, Meynet:15,Vink:23, Gonzalez-Tora:25}.  However, these approaches do not provide a single, physically motivated treatment that self-consistently captures eruptive mass loss in a stellar evolution program.

Recently, \cite{Cheng:24} implemented a physically-motivated prescription of super-Eddington mass loss in the \texttt{MESA} stellar evolution code.  In that work, a single efficiency parameter, $\xi$, governs the strength of radiation-driven winds when the energy associated with super-Eddington excess luminosity exceeds the gravitational binding energy. Their models demonstrated that including such enhanced mass loss during super-Eddington phases can significantly affect the evolution of RSGs. However, the efficiency parameter $\xi$ remains a critical free parameter that requires empirical calibration.

The Local Group of galaxies provides an ideal laboratory for constraining RSG mass-loss physics through detailed population studies. The proximity of the Large and Small Magellanic Clouds (LMC and SMC) and M31 allows for complete censuses of RSG populations with well-characterized distances, metallicities, and star formation histories \citep{Massey:03}. Additionally, these galaxies span a significant range in metallicity, from $\sim0.2~Z_\odot$ in the SMC to $\sim0.4~Z_\odot$ in the LMC and $\sim1.0~Z_\odot$ in M31 \citep{Russell:92, Sanders:12, Dalcanton:12, Choudhury:16}, enabling tests of potential metallicity-dependence in mass-loss prescriptions. Furthermore, each galaxy has been the subject of extensive photometric and spectroscopic surveys that have yielded large, well-characterized samples of confirmed RSGs suitable for statistical analysis.

In this work, we leverage RSG datasets from Local Group galaxies to constrain the efficiency parameter $\xi$ in the super-Eddington eruptive mass-loss prescription developed by \cite{Cheng:24}. We begin with selecting appropriate datasets featuring RSGs in the SMC, LMC, and M31 (Section \ref{sec:data}). Next, we generate grids of stellar evolution models using the \cite{Cheng:24} implementation of eruptive mass loss in the 1D stellar evolution software \texttt{MESA}, spanning the relevant range of initial masses (between $2-90~M_\odot$) and metallicities (at $0.2~Z_\odot$, $0.4~Z_\odot$, and $1~Z_\odot$), varying the efficiency parameter $\xi$ and metallicity $Z$ systematically to explore its effects on predicted RSG populations (Section \ref{sec:methods}). We then compare the resulting theoretical cumulative luminosity functions with the observed RSG censuses from the SMC, LMC, and M31, and derive constraints on $\xi$ (Section \ref{sec:results}). Finally, we summarize our results and place them in context in Section \ref{sec:discussion}.

\section{Data}
\label{sec:data}

In this section we describe the observational samples of the SMC, LMC, and M31 that are used to calibrate our eruptive mass loss prescription.

\subsection{SMC}

For the SMC we adopt the sample of RSGs from \cite{Yang:23}. Their sample is constructed by cross-matching Gaia-constrained RSG candidate lists from \cite{Yang:20} (1239 objects) and \cite{Ren:21} (2138 objects), removing duplicates for 2332 unique targets, and assembling photometry in 53 bands from the UV to the mid-IR (GALEX, SkyMapper, M2002, MCPS, NSC, Gaia, OGLE, DENIS, VMC, 2MASS/IRSF, AKARI, WISE, Spitzer). To improve data quality, they adopt SAGE--SMC in place of SEIP (Spitzer Enhanced Imaging Products), remove blended sources via the SAGE close-source flag, update WISE to \textit{AllWISE} with band-specific S/N thresholds, and refresh several surveys to their latest releases (Gaia EDR3, SkyMapper DR2, NSC DR2, VMC DR5). Additional cleaning addresses saturation and PSF issues and excises sources in the AGB--RSG overlap on the 2MASS CMD to enhance sample purity.  Foreground extinction is corrected with \(R_V=3.1\) and \(E(B{-}V)=0.033\), and bolometric luminosities are obtained by integrating the dereddened SEDs. They estimate effective temperatures by applying their adopted color-temperature calibration to extinction-corrected 2MASS $(J-K_S)$ color (with a total extinction correction corresponding to $A_V=0.1$ mag); their sample spans $\log_{10}(T_{\rm eff})\approx 3.5$--$3.75$. After these steps, the final SMC RSG sample contains 2121 stars and provides a homogeneous, quality-screened census for our comparison of theoretical predictions to observations.

\begin{figure*}[t!]
	\gridline{\fig{{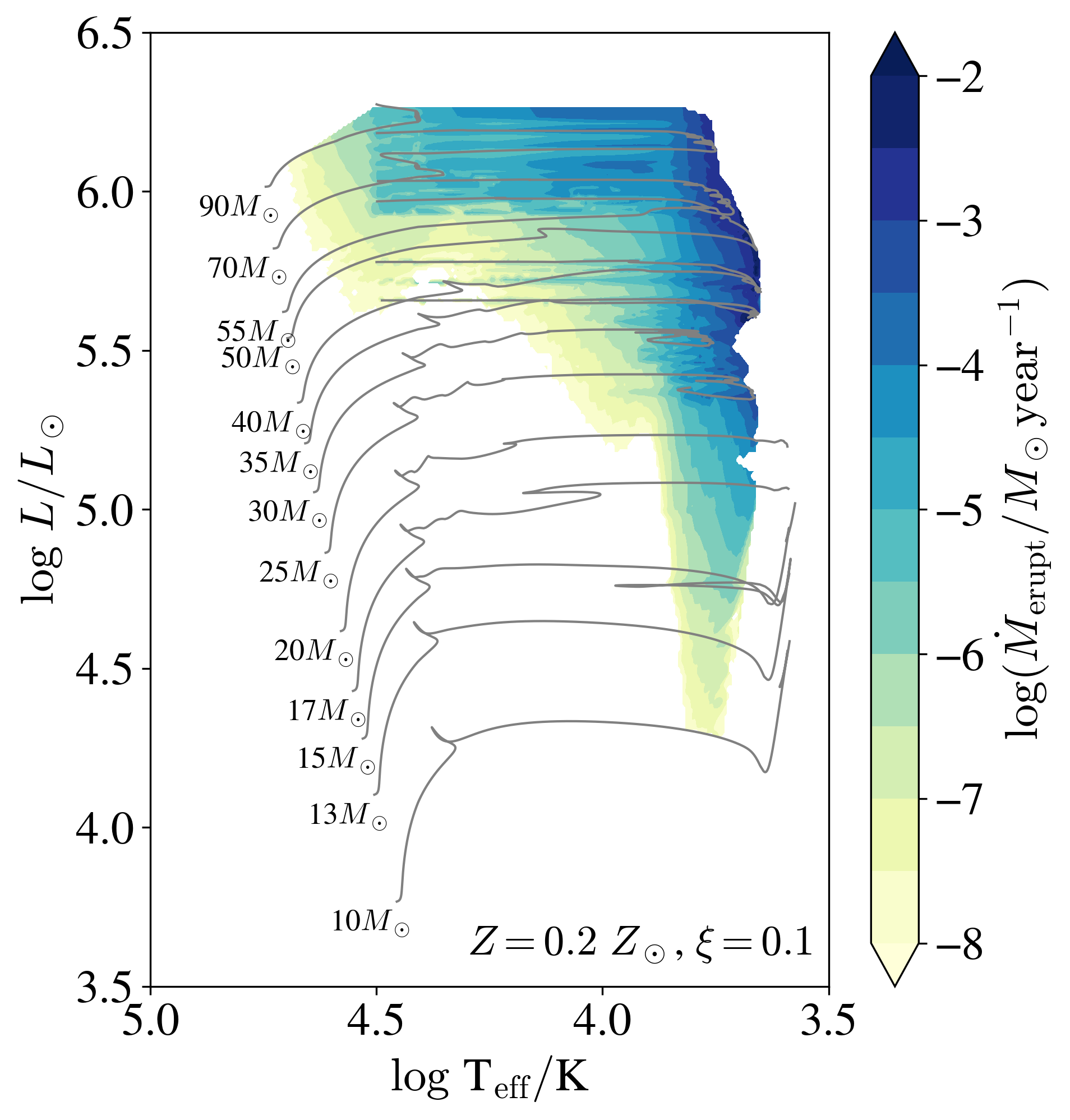}}{0.48\textwidth}{}
		\fig{{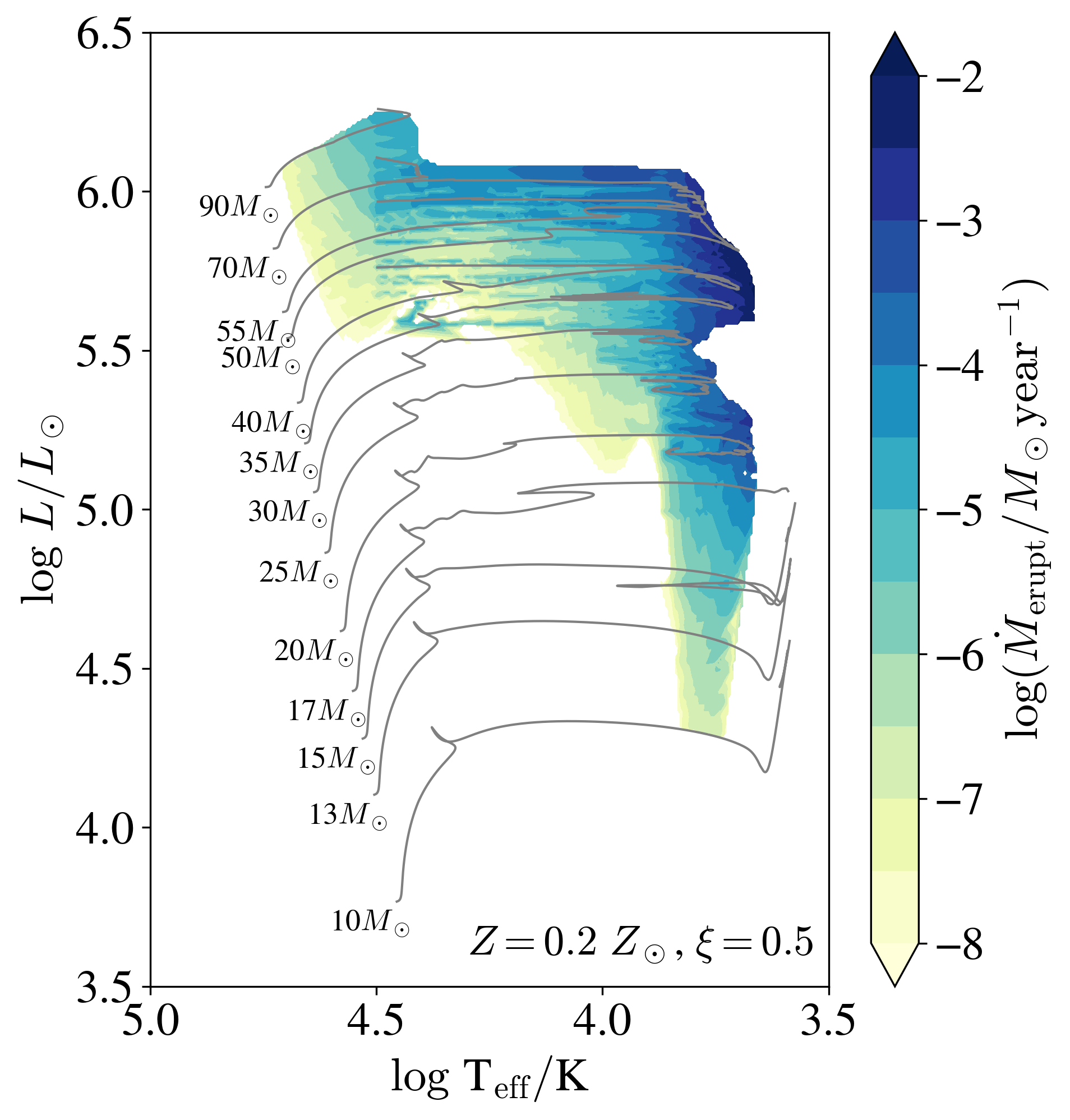}}{0.48\textwidth}{}
	}
    \vspace{-10pt}
	\caption{Eruptive mass loss rates across the H-R diagram for $Z=0.2~Z_\odot$. Left: $\xi=0.1$. Right: $\xi=0.5$. Grey lines show the evolutionary tracks of stars from the beginning of main sequence until the end of the model run. Models were run at every $1~M_\odot$ increments but only selected models are shown here for clarity. Colored contours show the rate of eruptive mass loss experienced by the star, with legend on the right.}\label{fig:SMC}
\end{figure*}

\begin{figure*}
\vspace{30pt}
	\gridline{\fig{{R1_LMC_1}.png}{0.48\textwidth}{}
		\fig{{R1_LMC_5}.png}{0.48\textwidth}{}}
	\vspace{-10pt}
	\caption{Eruptive mass loss rates across the H-R diagram for $Z=0.4~Z_\odot$. Left: $\xi=0.1$. Right: $\xi=0.5$. All axes, labels, and other lines follow Figure~\ref{fig:SMC}.}\label{fig:LMC} \vspace{6pt}
\end{figure*}

\begin{figure*}
\vspace{20pt}
	\gridline{\fig{{R1_M31_1}.png}{0.48\textwidth}{}
		\fig{{R1_M31_5}.png}{0.48\textwidth}{}}
	\vspace{-10pt}
	\caption{Eruptive mass loss rates across the H-R diagram for $Z=1.0~Z_\odot$. Left: $\xi=0.1$. Right: $\xi=0.5$. All axes, labels, and other lines follow Figure~\ref{fig:SMC}.}\label{fig:M31} 
\end{figure*}

\subsection{LMC}

For the LMC we adopt the RSG sample from \citet{Neugent:12} and restrict our analysis to their secure LMC members (``category~1''). Their supergiant candidates were initially selected from UCAC3, the USNO CCD Astrograph Catalog, an all-sky astrometric catalog that provides stellar positions, proper motions, and source-quality flags. They selected sources within the LMC's visible disk (a $3\fdg5$ radius around the LMC center), and applied a proper-motion cut that excludes objects with absolute proper motions $>15~\mathrm{mas\,yr^{-1}}$ in $\alpha$ or $\delta$ and used UCAC3 quality flags to remove likely non-stellar sources. Candidate RSGs were then selected in the 2MASS color--magnitude diagram; the RSG locus begins at $(J{-}K)_{\rm 2MASS}>0.9$ at $K_{\rm 2MASS}=10.2$, and they expect most RSGs to fall in $0.9<(J{-}K)_{\rm 2MASS}<1.2$ (corresponding to $T_{\rm eff}\sim3500$--$4500~\mathrm{K}$). LMC membership was established from multi-object spectroscopy and radial velocities; stars with $v_r>200~\mathrm{km\,s^{-1}}$ are classified as probable LMC supergiants (category~1), yielding 505 RSGs. For our work, we use the published $\log T_{\rm eff}$ and $\log(L/L_\odot)$ values tabulated for these category~1 RSGs (their Table~4), where $T_{\rm eff}$ is derived from de-reddened 2MASS $J{-}K$ colors (after conversion to the \citealt{Bessell:88} system and adopting $E(B{-}V)=0.13$) using MARCS-based transformations for RSGs, and bolometric luminosities are computed from the (transformed and de-reddened) $K$-band magnitudes using a $K$-band bolometric correction as a function of $T_{\rm eff}$.

\subsection{M31}

For M31 we use the RSG catalog of \citet{Massey:21}. Their catalog is based on near-IR $J$ and $K_s$ photometry over the M31 disk, with foreground stars removed using Gaia DR2 parallaxes and proper motions via a membership analysis. RSG candidates are then isolated in the $(J{-}K_s,\,K_s)$ color--magnitude diagram to minimize AGB contamination, followed by additional manual cleaning to remove obvious non-RSGs and spurious sources. They convert the NIR photometry to physical parameters using their tabulated transformations: extinction is treated with their adopted $A_V$ prescription, effective temperatures are derived from the intrinsic $(J{-}K)_0$ color, and bolometric luminosities are computed from the dereddened $K$ magnitude with a $K$-band bolometric correction as a function of $T_{\rm eff}$. For our work, we take the cataloged $T_{\rm eff}$ and $\log(L/L_\odot)$ values directly from their Table~3 and do not recompute reddening or bolometric corrections.

\section{Methods}
\label{sec:methods}

\subsection{Stellar Evolution Calculations with \texttt{MESA}}\label{sec:improve}

We evolved stellar models with Zero-Age Main Sequence (ZAMS) masses of $2-90~M_\odot$ using the 1D open-source stellar evolution software Modules for Experiments in Stellar Astrophysics (\texttt{MESA}; \citealt{Paxton:11,paxton:13,paxton:15,paxton:18,paxton:19,Jermyn:23}). To improve numerical robustness in the red–supergiant (RSG) regime, we updated our inlists compared to \citet{Cheng:24} for the \emph{baseline} (non-eruptive) mass-loss calculations to more closely follow the mesh and timestep controls adopted in \S7.2 of MESA VI \citep{Jermyn:23}. These changes were limited to solver tolerances and spatial/temporal resolution controls; all other input physics (opacities, EOS, nuclear network, convection, rotation, and steady winds) were left unchanged relative to the setup in \citet{Cheng:24}. Throughout, we use the same wind prescriptions and microphysics adopted in \citet{Cheng:24}. Inlists are accessible here at \dataset[doi: zenodo.19535148]{https://doi.org/10.5281/zenodo.19535148}.

To test metallicity scalings in the eruptive channel, we repeated the model grid at three representative metallicities,
\[
Z/Z_\odot \in \{0.2,\,0.4,\,1.0\},
\]
which we associate, respectively, with the SMC, LMC, and M31 for the purposes of comparison to Local Group populations. These values are chosen to match the metallicity baselines adopted in our introduction and are used consistently in all comparisons below.

\subsection{Mass Loss}
We consider two sources of mass-loss: a line-driven wind component and a super-Eddington, eruptive component:
\begin{equation}
	\dot{M} \;=\; \dot{M}_{\rm wind} \;+\; \dot{M}_{\rm erupt},
\end{equation}
where $\dot{M}$ is the total mass-loss rate, $\dot{M}_{\rm wind}$ is the standard steady-state wind contribution, and $\dot{M}_{\rm erupt}$ is the eruptive (super-Eddington) contribution.  In the following sections we describe our approach to these two components. 
\newpage
\subsubsection{Steady-state winds}

Our steady-state wind prescription is a standard hybrid approach.  At $T_{\rm eff}>4000~\mathrm{K}$ we use \texttt{MESA}'s \texttt{Dutch} prescription (\citealt{Vink:2001, Nugis:2000} as summarized by \citealt{Glebbeek:2009}) for $\dot{M}_{\rm wind}$, scaled by a fixed \texttt{Dutch\_scaling\_factor} of $0.8$ \citep[e.g.,][]{Glebbeek:2009,Fields:21,Johnston:24,Pereira:24,Shiber:24}. For $T_{\rm eff}<4000~\mathrm{K}$ we replace the default de Jager rates with the empirical RSG rates of \citet{Decin:2024}. This replaces the \citet{deJager:88} or \citet{Nieuwenhuijzen:90} mass loss rates typically used in \texttt{MESA}'s `\texttt{Dutch}' prescription which are considered to overestimate mass loss rates \citep[e.g.,][]{Beasor:20,Beasor:2023,Antoniadis:24}. The choice of steady-state wind model at cool temperatures does not significantly affect the evolution of models with strong eruptive winds, since such models do not evolve cool enough to enter the \citet{Decin:2024} regime. 

\subsubsection{Eruptive winds}

Our eruptive wind model was described in detail in \citet{Cheng:24}; we provide only a summary here.  Eruptive mass loss is triggered where the local radiative luminosity exceeds the Eddington limit in layers that are optically coupled but have inefficient convection. We adopt
\begin{equation}
	\tau_{\rm crit} \equiv \frac{c}{c_s}, \qquad 
	L_{\rm Edd} \equiv \frac{4\pi G c\,M_{\rm enc}}{\kappa},
\end{equation}
where $c$ is the speed of light, $c_s$ is the local sound speed, $\tau$ denotes the local optical depth, and $\tau_{\rm crit}$ is the `critical' optical depth outside of which the photon diffusion speed becomes comparable to the local sound speed. $L_{\rm Edd}$ is the local Eddington luminosity, $G$ is the gravitational constant, $M_{\rm enc}$ is the enclosed mass at radius $r$, and $\kappa$ is the local opacity. We consider zones with $2/3 \lesssim \tau < \tau_{\rm crit}$ and $L_{\rm rad}>L_{\rm Edd}$, where $L_{\rm rad}$ is the local radiative luminosity. The excess radiative luminosity
\begin{equation}
	L_{\rm excess} \equiv \max\!\left[\,0,\,L_{\rm rad}-L_{\rm Edd}\,\right]
\end{equation}
is converted to a local excess energy over one local dynamical time,
\begin{equation}
	t_{\rm dyn}(r)=\sqrt{\frac{r^{3}}{G\,m(r)}}, \qquad
	E_{\rm excess}(r)=L_{\rm excess}\,t_{\rm dyn}(r),
\end{equation}
where $m(r)$ is the enclosed mass at radius $r$; $E_{\rm excess}(r)$ is the excess energy available over one dynamical time.

From the base of the region ($\tau=\tau_{\rm crit}$) outward we form a cumulative excess energy,
\begin{equation}
	E_{\rm tot,excess}(r)=
	\frac{\displaystyle\int_{m[r(\tau_{\rm crit})]}^{m(r)} E_{\rm excess}\,{\rm d}m}
	{\displaystyle\int_{m[r(\tau_{\rm crit})]}^{m(r)} {\rm d}m},
\end{equation}
where $r(\tau_{\rm crit})$ is the radius at which $\tau=\tau_{\rm crit}$.

We compare this to the gravitational binding energy of the overlying envelope,
\begin{equation}
	E_{\rm bind}(r)=-\int_{m(r)}^{M}\frac{G\,m}{r}\,{\rm d}m,
\end{equation}
where $M$ is the total stellar mass.

We locate the smallest radius above $r(\tau_{\rm crit})$ where $E_{\rm tot,excess}(r)>\lvert E_{\rm bind}(r)\rvert$ and define
\begin{equation}
	\Delta M \equiv M - m\!\left(r_{\rm mass\;loss}\right), \qquad
	\dot{M}_{\rm erupt} \equiv \xi\,\frac{\Delta M}{t_{\rm dyn}\!\left(r_{\rm mass\;loss}\right)} ,
\end{equation}
where $r_{\rm mass\;loss}$ is the radius where the above condition is first satisfied, $\Delta M$ is the envelope mass exterior to $r_{\rm mass\;loss}$, and $\xi$ is a dimensionless efficiency (free) parameter calibrated against observations.

An explicit metallicity dependence is not imposed in $\dot{M}_{\rm erupt}$; any $Z$ (metallicity) trends arise indirectly through structural changes (e.g., $T_{\rm eff}$ and opacity peaks). See \cite{Cheng:24} for the full derivation and implementation details.

The overall strength of the eruptive model is determined by the dimensionless efficiency parameter $\xi$. Our goal in this work is to leverage observational data to calibrate $\xi$. Thus, we computed models for a grid
\[
\xi \in \{0.0,\, 0.05,\,0.1,\,0.2,\,0.3,\,0.4,\,0.5\},
\]
where $\xi=0.0$ corresponds to a control case with only the steady (non-eruptive) winds active. For the $Z=0.2\ Z_\odot$ models we ran an additional grid with $\xi = 0.025$ for comparison to observations.

The \texttt{MESA} models and eruptive mass loss rates for  $Z=0.2~Z_\odot, 0.4~Z_\odot, 1.0~Z_\odot$ are shown in Figures~\ref{fig:SMC}, \ref{fig:LMC}, and \ref{fig:M31}, respectively.  Note that these models evolved further compared to those shown in \cite{Cheng:24} due to the improved baseline \texttt{MESA} models (see Section~\ref{sec:improve}). Stars with $M<50M_\odot$ evolve through core helium depletion. At higher masses, stars evolve past core helium ignition but many do not reach the end of core helium burning due to numerical issues (see Section~$2$ of \cite{Cheng:24} for more details). We additionally exclude a small number of individual tracks that show clear outlier behavior consistent with numerical artifacts when plotted in the H–R diagram (e.g., isolated, discontinuous excursions not seen in the surrounding grid). For our purposes of comparison with observations, this level of model completion is sufficient, as all modeled stars of $50~M_\odot$ and below evolved far enough to probe the RSG region of the Hertzsprung-Russell (H-R) diagram.

\subsection{Population Synthesis and Mock Observables}

To connect our evolutionary models to observable distributions, we convert our custom \texttt{MESA} tracks into EEP (equivalent evolutionary points) form and generate \texttt{MIST}-style isochrones using the EEP framework \citep[e.g.,][]{Dotter:16,Choi:16}. We then synthesize mock stellar populations from these isochrones with \texttt{ArtPop} \citep{Greco:22}, sampling initial masses from a \citet{Kroupa:01} IMF and ages from a constant star formation rate. The SFR adopted for SMC, LMC, and M31 are $0.052~M_\odot~\mathrm{yr^{-1}}$ \citep{Zapartas:25}, $0.4~M_\odot~\mathrm{yr^{-1}}$ \citep{Harris:09}, and $0.35~M_\odot~\mathrm{yr^{-1}}$ \citep{Rahmani:16}, respectively (although we note that the absolute value of the SFR does not impact our comparisons to observations as we normalize the overall model counts to the data). We record $\log(T_{\rm eff}/\mathrm{K})$ and $\log(L/L_\odot)$ for H-R diagrams and cumulative luminosity-function comparisons; photometric transformations in \texttt{ArtPop} are used only for visualization. To approximate uncertainties in the observational literature sources, we add Gaussian noise with standard deviations of $0.03$ dex in $\log(T_{\rm eff}/\mathrm{K})$ and $0.1$ dex in $\log(L/L_\odot)$.

\begin{figure*}[ht!]
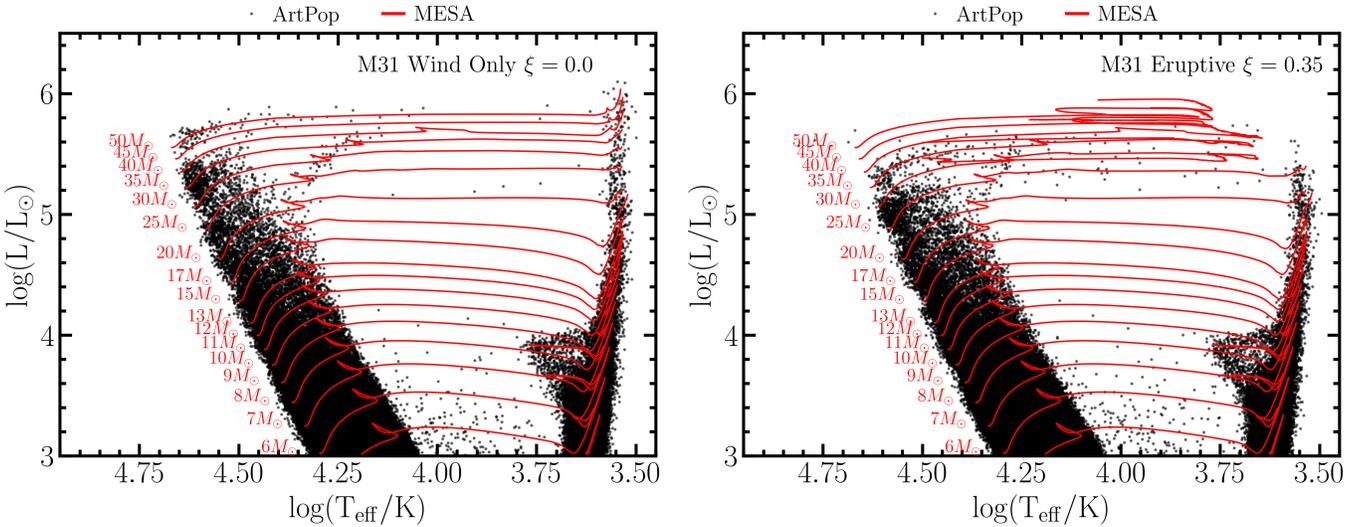

\gridline{\fig{{2_tracks+theory+control_alt}.png}{0.5\textwidth}{}
		\fig{{2_tracks+theory_alt2}.png}{0.5\textwidth}{}}
	\vspace{-20pt}
	\caption{H-R diagram for $Z = 1.0~Z_\odot$. Left: $\xi=0.0$. Right: $\xi=0.35$. \texttt{MESA} tracks are in red lines and \texttt{ArtPop}-simulated population generated from the \texttt{MESA} model grid are in black dots. For clarity, only models from $6-50~M_\odot$ are shown. Gaussian noise of $0.1~\mathrm{dex}$ and $0.03~\mathrm{dex}$ were added for $\log(\mathrm{L/L_\odot})$ and $\log(\mathrm{T_{\rm eff}/K})$ respectively to mimic observational scatter.}\label{fig:HR}
\end{figure*}

\subsection{Luminosity Functions}

To obtain luminosity functions useful for comparison, we start with an H-R diagram similar to those shown in Figure~\ref{fig:HR}. Next, we select stars in the range to $3.5 < \log T_{\rm eff} < 3.75$ and $\log(L/L_\odot) > 4.5$ so as to focus on the RSG region of the H-R diagram. Finally, we calculate the cumulative luminosity function, integrating from high to low $\log(L/L_\odot)$ and normalizing the total number of the \texttt{ArtPop} output stars to the total number of observed stars within the $\log T_{\rm eff}$ and $\log(L/L_\odot)$ bounds. We use the cumulative luminosity function, i.e., the number of stars brighter than a given luminosity, because it reduces binning sensitivity and suppresses small number noise while directly highlighting differences in the bright tail.
\begin{figure*}[t!]
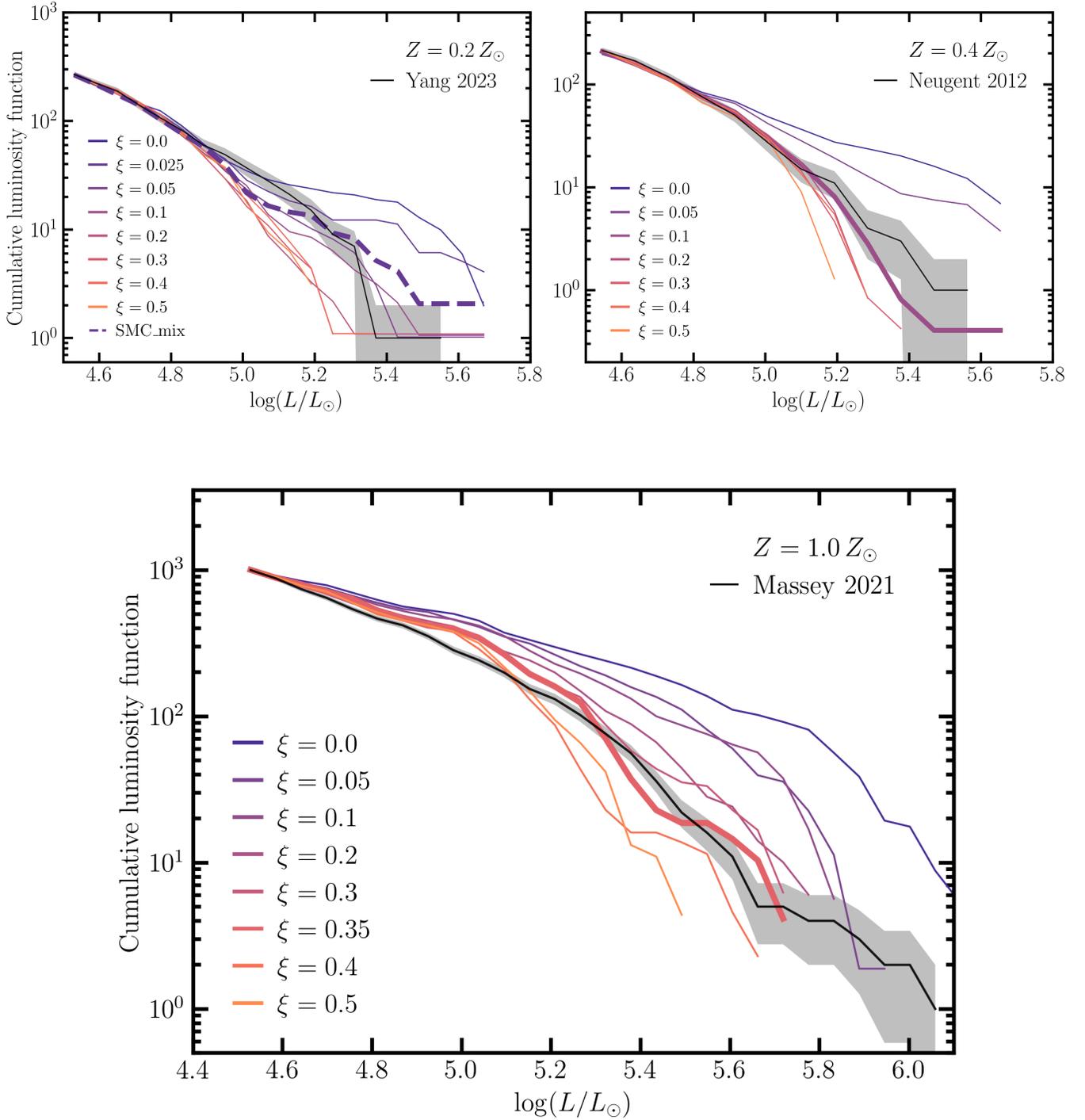

	\centering
    \vspace{50pt}
	\gridline{\fig{{LF_SMC_LMC_2paneltest}.png}{1.0\textwidth}{}\vspace{0pt}}
	\gridline{\fig{{LF_M31_35}.png}{0.8\textwidth}{}}
	\caption{Cumulative luminosity functions for each modeled stellar population compared to observations of the SMC (upper left panel, $Z=0.2~Z_\odot$), LMC (upper right panel, $Z=0.4~Z_\odot$), and M31 ($Z=1.0~Z_\odot$, lower panel). Shown are both cumulative luminosity functions from \texttt{ArtPop} population synthesis output (colored lines) and observations (black lines). The thicker colored line (either dashed or solid) indicates the best fit model. Grey shading is the Poisson error of the observed data.}\label{fig:res}
\end{figure*}
\subsection{Convergence and Robustness}

We verified that all reported trends with $\xi$ and $Z$ are insensitive to reasonable tightening of the spatial and temporal controls in our inlists. Specifically, we re-ran representative models with stricter timestep and mesh controls (following the guidance in \S7.2 of \citealt{Jermyn:23}) and confirmed that the resulting RSG lifetimes and luminosity distributions used in our analysis are numerically robust. Because the eruptive prescription is unchanged from \citet{Cheng:24}, this procedure isolates the impact of $\xi$ and $Z$ from purely numerical effects.

\begin{figure}[ht!]
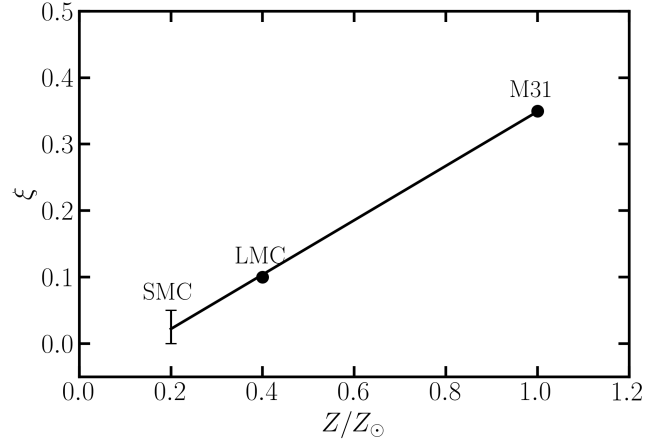

    \vspace{10pt}  
	\fig{{xi_vs_metallicity_35}.png}{0.48\textwidth}{}
	\vspace{-20pt}
	\caption{Preferred $\xi$ values based on comparison of observed and modeled RSG luminosity functions (see Figure~\ref{fig:res}). The black line is a linear fit to the data points, with equation $\xi = 0.40\,(Z/Z_\odot) - 0.06$. For the SMC, the linear fit was performed with the midpoint of the best fit range.}\label{fig:final}
\end{figure}

\section{Results} \label{sec:results}

In this section we present and discuss the results from our \texttt{MESA}-derived mock observables and our determination of the best fit $\xi$ free parameters at different metallicities. 

We start with \texttt{MESA} stellar evolution tracks, which we then transform into a realistic stellar population with an appropriate star formation history (using the \texttt{MIST} procedure combined with \texttt{ArtPop}). This is shown in Figure~\ref{fig:HR}, which compares the \texttt{MESA} tracks against the modeled stellar population. We show two $Z$ and $\xi$ combinations ($Z = 1.0~Z_\odot$ with $\xi = 0.0$ and $\xi = 0.35$). While \texttt{MESA} tracks are not shown for every solar mass increment in Figure~\ref{fig:HR} (for clarity and visibility), the stellar population was generated using the full sample of \texttt{MESA} tracks that evolved fully to core helium depletion (from $2-50~M_\odot$ in increments of $1~M_\odot$). Nevertheless, in Figure~\ref{fig:HR}, the stellar population clearly follows the \texttt{MESA} evolutionary tracks, though not perfectly since a small amount of Gaussian random noise was applied to approximate observational uncertainties. As expected, stars are found in higher numbers on the main sequence as well as on the RSG branch - long-lived phases in the evolution. Comparing the two panels, we can see that the presence of eruptive mass loss (i.e. non-zero $\xi$) reduces the number of RSGs in the population.

Figure~\ref{fig:res} contains our main results. We show luminosity functions of SMC, LMC, and M31 at varying $\xi$ and compare against their respective observational samples: \cite{Yang:23} for SMC, \cite{Neugent:12} for LMC, and \cite{Massey:21} for M31.  For the SMC and LMC, the data and models agree well at lower luminosities (log $L/L\odot\lesssim5$), with or without the additional eruptive mass-loss prescription.  At higher luminosities the data clearly favor a non-zero value for the eruptive efficiency parameter, $\xi$.  For M31 there is moderate tension between the data and models even at lower luminosities.  The origin of the tension is unclear, but could be due to challenges in contamination or completeness in the observations, or one or more of our assumptions on the modeling side (including the assumed constant star formation history for M31).  In any event, the tension between data and models becomes much larger at higher luminosities unless an eruptive mass-loss model is included. 

Based on these cumulative luminosity functions, our preferred values for $\xi$ are:
\begin{align*}
\xi_\mathrm{SMC} &= \begin{cases}
0.0 \quad & \text{for} \ M \lesssim 15~M_\odot \ (\log(L/L_\odot) \lesssim 5.0),\\
0.05 \quad & \text{for}  \ M \gtrsim 15~M_\odot \ (\log(L/L_\odot) \gtrsim 5.0)~.
\end{cases}\\
\xi_\mathrm{LMC} &= 0.10\\
\xi_\mathrm{M31} &= 0.35 \ \mathrm{.}
\end{align*}
where $M$ is the initial stellar mass. We estimate $\xi_{\rm best}$ via a ``$\chi$-by-eye'' comparison of the cumulative luminosity functions. We considered adopting a formal likelihood/metric, but because the model and observed distributions show systematic shape offsets (i.e., no $\xi$ provides a uniformly good match across the full luminosity range), a single goodness-of-fit statistic would mainly quantify the overall mismatch rather than robustly distinguish among neighboring $\xi$ values; we therefore use the cumulative comparison to select the $\xi$ that most nearly reproduces the observed distribution shape. If instead we adopt a supersolar metallicity for M31 ($Z=1.5\,Z_\odot$), the best fit value shifts to $\xi_{\rm best,M31}\simeq 0.7$; this case is not our fiducial choice because the super-solar metallicity models are less numerically robust in 
\texttt{MESA}. We note that the observational sample used in the M31 calibration, namely \citet{Massey:21}, includes “a dozen M31 stars [that] have unlikely large extinction values ($A_V = 5 - 10$) and unrealistically high luminosities.” If a ceiling of $\log(L/L_\odot) = 5.5$ is imposed, the preferred value for $\xi$ for M31 is $0.35$ for $\log(L/L_\odot) < 5.4$, though $\xi = 0.5$ produces an overall better fit across the whole $\log(L/L_\odot)$ range.

Figure~\ref{fig:final} summarizes the best fit $\xi$ values, and suggests a monotonic increase of $\xi$ with metallicity, though with only three systems we treat this trend as suggestive. A discussion of the possible causes of this trend as well as directions for future work is included in Section~\ref{sec:discussion}. 

\newpage
\section{Discussion and Conclusions} \label{sec:discussion}

This work uses resolved RSG populations in the SMC, LMC, and M31 to calibrate the eruptive mass-loss efficiency parameter $\xi$ introduced in \cite{Cheng:24}. We found that the best fit $\xi$ values across the three systems are: $\xi_{\rm best,SMC}\simeq0.0-0.05$, $\xi_{\rm best,LMC}\simeq0.10$, and $\xi_{\rm best,M31}\simeq0.35$. With only three galaxies studied, we treat the monotonically increasing $\xi-Z$ relationship as suggestive. 

To obtain these constraints, we evolved stellar models with \texttt{MESA} using the same baseline winds and microphysics as in \citet{Cheng:24}, and implemented their super-Eddington eruptive mass loss model parameterized by $\xi$. We then converted the resulting tracks to EEP-based, \texttt{MIST}-style isochrones and synthesized mock populations with \texttt{ArtPop} for comparison to observed samples \citep{Yang:23,Neugent:12,Massey:21} via cumulative luminosity functions. We included initial masses extending to $90\,M_\odot$, with all models at $M_{\rm init}\le 50\,M_\odot$ evolved consistently through core helium depletion (and higher mass models evolved at least past core helium ignition), which is sufficient for our purposes. 

We emphasize that the efficiency parameter $\xi$ calibrated in this work governs only the eruptive mass loss channel, and leaves the steady-wind prescription unchanged. This is a deliberate choice. Several groups have previously addressed the observed deficit of luminous red supergiants by enhancing steady wind mass loss rates, in some cases by factors of $10-25$ or through $\Gamma$-dependent prescriptions motivated by observed mass loss features \citep[e.g.,][]{Ekstrom:12, Meynet:15, Vink:23}. There is broad agreement that the physical origin of this additional mass loss is connected to stars approaching the Eddington limit, and the eruptive prescription of \citet{Cheng:24} models this from first principles, self-consistently triggering mass loss when the local super-Eddington excess energy exceeds the envelope binding energy. Both approaches aim to capture the same underlying near-Eddington driven mass loss, and a systematic comparison of their respective predictions is an interesting avenue for future work.

The origin of the tentative relation between $\xi$ and metallicity is intriguing.  One possibility for the origin of this trend is that $\xi$ is capturing not only what triggers an eruption, but how efficiently an eruption turns into net mass loss. Conceptually, that efficiency could be set at two stages: the launching of material out of the deep envelope, and the subsequent propulsion needed for that material to actually escape rather than fall back. There is some observational evidence that RSG envelope convection itself varies with metallicity: observational calibrations of RSG temperatures suggest a metallicity-dependent mixing length, which \citet{Chun:18} interpret as more efficient convective energy transport at higher $Z$. Models rerun with these metallicity-calibrated \texttt{alpha\_MLT} values produced only minor changes, which are absorbed into $\xi$. Other metallicity-dependent differences in envelope convective structure could in principle also contribute to the inferred $\xi-Z$ trend. Such effects are not explicitly separated in our models and would be absorbed into $\xi$ alongside the eruptive mass-loss efficiency. Disentangling these contributions would require varying convective treatments across metallicities, which we leave to future work. A first-principles treatment of convective effects on stellar structure would ultimately require 3D simulations.

On the launching side, the relevant balance is between the energetics of the envelope (e.g., convective driving and other sources of excess energy) and the degree to which radiation can couple to and do work on the gas. At higher metallicity, increased opacity (including metal line opacity as well as continuum features such as the iron bump) could plausibly improve that coupling, making it easier for a given energy injection to lift material to large radii \citep{CAK:75, Iglesias1993, Jiang:15, Jiang:18}.  On the escape side, higher-$Z$ environments may also provide additional help in accelerating the lifted material outward, for example through opacity sources in cool outflows \citep[including dust, e.g.][]{vanLoon:05,Goldman:17}, which would again increase the fraction of launched mass that ultimately becomes unbound. In this picture, $\xi$ may be absorbing metallicity-dependent physics in either or both of the launching and escape stages.

We note that the three observational samples used in this work employ different methodologies. Luminosities in the SMC sample \citep{Yang:23} are derived from full SED integration across 53 photometric bands, while those in the LMC \citep{Neugent:12} and M31 \citep{Massey:21} samples rely on K-band bolometric corrections as a function of effective temperature. The extinction treatments also differ: \citet{Yang:23} apply a uniform foreground correction of $E(B-V) = 0.033$ for the SMC, \citet{Neugent:12} adopt a uniform $E(B-V) = 0.13$ for the LMC, while \citet{Massey:21} use a luminosity-dependent extinction for M31. Such methodological differences could introduce systematic offsets in the luminosity scales across the three galaxies, which would shift their luminosity functions relative to one another and potentially bias the inferred $\xi$ values. This reinforces our conclusion that the $\xi-Z$ trend should be treated as suggestive rather than definitive and serves as motivation for future investigation.

There are several directions for future work.  On the observational side, expanding the comparison beyond three systems and exploring a finer and broader grid of $Z$ would test whether the $\xi-Z$ trend is robust and if our general eruptive mass-loss scheme is able to capture the detailed shape of the RSG luminosity function. On the modeling side, detailed three-dimensional radiation hydrodynamic simulations of eruptive mass loss at multiple metallicities and including the effects of dust would help separate the physics of mass launching from escape and determine whether metallicity primarily affects the trigger, the escape efficiency, or both.

\section{Data Availability}

SMC observational data were obtained from a VizieR Online Data Catalog: \citet{10.26093/cds/vizier.36760084}.

For the LMC and M31, observational data were obtained from Table 4 of \citet{Neugent:12} and Table 3 of \citet{Massey:21} respectively, specifically the online journal article version.

\section{Acknowledgments}
 We thank Matteo Cantiello, Evan B. Bauer, and Mathieu Renzo for helpful conversations surrounding the modelling of eruptive mass loss. S.J.C. and C.C. acknowledge support from NSF grant AST2009131. J.A.G. acknowledges financial support from NASA grant 23-ATP23-0070. The Flatiron Institute is supported by the Simons Foundation. 

\newpage
\bibliography{references}{}

\begin{thebibliography}{}
\expandafter\ifx\csname natexlab\endcsname\relax\def\natexlab#1{#1}\fi
\providecommand{\url}[1]{\href{#1}{#1}}
\providecommand{\dodoi}[1]{doi:~\href{http://doi.org/#1}{\nolinkurl{#1}}}
\providecommand{\doeprint}[1]{\href{http://ascl.net/#1}{\nolinkurl{http://ascl.net/#1}}}
\providecommand{\doarXiv}[1]{\href{https://arxiv.org/abs/#1}{\nolinkurl{https://arxiv.org/abs/#1}}}

\bibitem[{{Antoniadis} {et~al.}(2024){Antoniadis}, {Bonanos}, {de Wit}, {Zapartas}, {Munoz-Sanchez}, \& {Maravelias}}]{Antoniadis:24}
{Antoniadis}, K., {Bonanos}, A.~Z., {de Wit}, S., {et~al.} 2024, \aap, 686, A88, \dodoi{10.1051/0004-6361/202449383}

\bibitem[{{Banerjee} {et~al.}(2020){Banerjee}, {Belczynski}, {Fryer}, {Berczik}, {Hurley}, {Spurzem}, \& {Wang}}]{Banerjee:20}
{Banerjee}, S., {Belczynski}, K., {Fryer}, C.~L., {et~al.} 2020, \aap, 639, A41, \dodoi{10.1051/0004-6361/201935332}

\bibitem[{{Beasor} {et~al.}(2020){Beasor}, {Davies}, {Smith}, {van Loon}, {Gehrz}, \& {Figer}}]{Beasor:20}
{Beasor}, E.~R., {Davies}, B., {Smith}, N., {et~al.} 2020, \mnras, 492, 5994, \dodoi{10.1093/mnras/staa255}

\bibitem[{{Beasor} {et~al.}(2023){Beasor}, {Davies}, {Smith}, {van Loon}, {Gehrz}, \& {Figer}}]{Beasor:2023}
---. 2023, \mnras, 524, 2460, \dodoi{10.1093/mnras/stad1818}

\bibitem[{{Belczynski} {et~al.}(2010){Belczynski}, {Bulik}, {Fryer}, {Ruiter}, {Valsecchi}, {Vink}, \& {Hurley}}]{Belczynski:10}
{Belczynski}, K., {Bulik}, T., {Fryer}, C.~L., {et~al.} 2010, \apj, 714, 1217, \dodoi{10.1088/0004-637X/714/2/1217}

\bibitem[{Bessell \& Brett(1988)}]{Bessell:88}
Bessell, M.~S., \& Brett, J.~M. 1988, Publications of the Astronomical Society of the Pacific, 100, 1134, \dodoi{10.1086/132281}

\bibitem[{{Bruch} {et~al.}(2023){Bruch}, {Gal-Yam}, {Yaron}, {Chen}, {Strotjohann}, {Irani}, {Zimmerman}, {Schulze}, {Yang}, {Kim}, {Bulla}, {Sollerman}, {Rigault}, {Ofek}, {Soumagnac}, {Masci}, {Fremling}, {Perley}, {Nordin}, {Cenko}, {Ho}, {Adams}, {Adreoni}, {Bellm}, {Blagorodnova}, {Burdge}, {De}, {Dekany}, {Dhawan}, {Drake}, {Duev}, {Graham}, {Graham}, {Jencson}, {Karamehmetoglu}, {Kasliwal}, {Kulkarni}, {Miller}, {Neill}, {Prince}, {Riddle}, {Rusholme}, {Sharma}, {Smith}, {Sravan}, {Taggart}, {Walters}, \& {Yan}}]{Bruch2023}
{Bruch}, R.~J., {Gal-Yam}, A., {Yaron}, O., {et~al.} 2023, \apj, 952, 119, \dodoi{10.3847/1538-4357/acd8be}

\bibitem[{{Castor} {et~al.}(1975){Castor}, {Abbott}, \& {Klein}}]{CAK:75}
{Castor}, J.~I., {Abbott}, D.~C., \& {Klein}, R.~I. 1975, \apj, 195, 157, \dodoi{10.1086/153315}

\bibitem[{{Ceverino} \& {Klypin}(2009)}]{Ceverino:09}
{Ceverino}, D., \& {Klypin}, A. 2009, \apj, 695, 292, \dodoi{10.1088/0004-637X/695/1/292}

\bibitem[{{Cheng} {et~al.}(2024){Cheng}, {Goldberg}, {Cantiello}, {Bauer}, {Renzo}, \& {Conroy}}]{Cheng:24}
{Cheng}, S.~J., {Goldberg}, J.~A., {Cantiello}, M., {et~al.} 2024, \apj, 974, 270, \dodoi{10.3847/1538-4357/ad701e}

\bibitem[{{Choi} {et~al.}(2016){Choi}, {Dotter}, {Conroy}, {Cantiello}, {Paxton}, \& {Johnson}}]{Choi:16}
{Choi}, J., {Dotter}, A., {Conroy}, C., {et~al.} 2016, \apj, 823, 102, \dodoi{10.3847/0004-637X/823/2/102}

\bibitem[{{Choudhury} {et~al.}(2016){Choudhury}, {Subramaniam}, \& {Cole}}]{Choudhury:16}
{Choudhury}, S., {Subramaniam}, A., \& {Cole}, A.~A. 2016, \mnras, 455, 1855, \dodoi{10.1093/mnras/stv2414}

\bibitem[{{Chun} {et~al.}(2018){Chun}, {Yoon}, {Jung}, {Kim}, \& {Kim}}]{Chun:18}
{Chun}, S.-H., {Yoon}, S.-C., {Jung}, M.-K., {Kim}, D.~U., \& {Kim}, J. 2018, \apj, 853, 79, \dodoi{10.3847/1538-4357/aa9a37}

\bibitem[{{Conroy} \& {Gunn}(2010)}]{Conroy:10}
{Conroy}, C., \& {Gunn}, J.~E. 2010, \apj, 712, 833, \dodoi{10.1088/0004-637X/712/2/833}

\bibitem[{{Dalcanton} {et~al.}(2012){Dalcanton}, {Williams}, {Lang}, {Lauer}, {Kalirai}, {Seth}, {Dolphin}, {Rosenfield}, {Weisz}, {Bell}, {Bianchi}, {Boyer}, {Caldwell}, {Dong}, {Dorman}, {Gilbert}, {Girardi}, {Gogarten}, {Gordon}, {Guhathakurta}, {Hodge}, {Holtzman}, {Johnson}, {Larsen}, {Lewis}, {Melbourne}, {Olsen}, {Rix}, {Rosema}, {Saha}, {Sarajedini}, {Skillman}, \& {Stanek}}]{Dalcanton:12}
{Dalcanton}, J.~J., {Williams}, B.~F., {Lang}, D., {et~al.} 2012, \apjs, 200, 18, \dodoi{10.1088/0067-0049/200/2/18}

\bibitem[{{de Jager} {et~al.}(1988){de Jager}, {Nieuwenhuijzen}, \& {van der Hucht}}]{deJager:88}
{de Jager}, C., {Nieuwenhuijzen}, H., \& {van der Hucht}, K.~A. 1988, \aaps, 72, 259

\bibitem[{{Decin} {et~al.}(2024){Decin}, {Richards}, {Marchant}, \& {Sana}}]{Decin:2024}
{Decin}, L., {Richards}, A.~M.~S., {Marchant}, P., \& {Sana}, H. 2024, \aap, 681, A17, \dodoi{10.1051/0004-6361/202244635}

\bibitem[{{Delbroek} {et~al.}(2025){Delbroek}, {Sundqvist}, {Debnath}, {Moens}, {Backs}, {Van der Sijpt}, {Verhamme}, \& {Schillemans}}]{Delbroek:2025}
{Delbroek}, L., {Sundqvist}, J.~O., {Debnath}, D., {et~al.} 2025, \aap, 704, A104, \dodoi{10.1051/0004-6361/202556421}

\bibitem[{{Dessart} {et~al.}(2013){Dessart}, {Hillier}, {Waldman}, \& {Livne}}]{Dessart:13}
{Dessart}, L., {Hillier}, D.~J., {Waldman}, R., \& {Livne}, E. 2013, \mnras, 433, 1745, \dodoi{10.1093/mnras/stt861}

\bibitem[{{Dominik} {et~al.}(2012){Dominik}, {Belczynski}, {Fryer}, {Holz}, {Berti}, {Bulik}, {Mandel}, \& {O'Shaughnessy}}]{Dominik:12}
{Dominik}, M., {Belczynski}, K., {Fryer}, C., {et~al.} 2012, \apj, 759, 52, \dodoi{10.1088/0004-637X/759/1/52}

\bibitem[{{Dotter}(2016)}]{Dotter:16}
{Dotter}, A. 2016, \apjs, 222, 8, \dodoi{10.3847/0067-0049/222/1/8}

\bibitem[{{Ekstr{\"o}m} {et~al.}(2012){Ekstr{\"o}m}, {Georgy}, {Eggenberger}, {Meynet}, {Mowlavi}, {Wyttenbach}, {Granada}, {Decressin}, {Hirschi}, {Frischknecht}, {Charbonnel}, \& {Maeder}}]{Ekstrom:12}
{Ekstr{\"o}m}, S., {Georgy}, C., {Eggenberger}, P., {et~al.} 2012, \aap, 537, A146, \dodoi{10.1051/0004-6361/201117751}

\bibitem[{{Fields} \& {Couch}(2021)}]{Fields:21}
{Fields}, C.~E., \& {Couch}, S.~M. 2021, \apj, 921, 28, \dodoi{10.3847/1538-4357/ac24fb}

\bibitem[{{Fragos} {et~al.}(2023){Fragos}, {Andrews}, {Bavera}, {Berry}, {Coughlin}, {Dotter}, {Giri}, {Kalogera}, {Katsaggelos}, {Kovlakas}, {Lalvani}, {Misra}, {Srivastava}, {Qin}, {Rocha}, {Rom{\'a}n-Garza}, {Serra}, {Stahle}, {Sun}, {Teng}, {Trajcevski}, {Tran}, {Xing}, {Zapartas}, \& {Zevin}}]{Fragos:23}
{Fragos}, T., {Andrews}, J.~J., {Bavera}, S.~S., {et~al.} 2023, \apjs, 264, 45, \dodoi{10.3847/1538-4365/ac90c1}

\bibitem[{{Geen} {et~al.}(2015){Geen}, {Rosdahl}, {Blaizot}, {Devriendt}, \& {Slyz}}]{Geen:15}
{Geen}, S., {Rosdahl}, J., {Blaizot}, J., {Devriendt}, J., \& {Slyz}, A. 2015, \mnras, 448, 3248, \dodoi{10.1093/mnras/stv251}

\bibitem[{{Glebbeek} {et~al.}(2009){Glebbeek}, {Gaburov}, {de Mink}, {Pols}, \& {Portegies Zwart}}]{Glebbeek:2009}
{Glebbeek}, E., {Gaburov}, E., {de Mink}, S.~E., {Pols}, O.~R., \& {Portegies Zwart}, S.~F. 2009, \aap, 497, 255, \dodoi{10.1051/0004-6361/200810425}

\bibitem[{{Goldman} {et~al.}(2017){Goldman}, {van Loon}, {Zijlstra}, {Green}, {Wood}, {Nanni}, {Imai}, {Whitelock}, {Matsuura}, {Groenewegen}, \& {G{\'o}mez}}]{Goldman:17}
{Goldman}, S.~R., {van Loon}, J.~T., {Zijlstra}, A.~A., {et~al.} 2017, \mnras, 465, 403, \dodoi{10.1093/mnras/stw2708}

\bibitem[{{Gonz{\'a}lez-Tor{\`a}} {et~al.}(2025){Gonz{\'a}lez-Tor{\`a}}, {Sander}, {Sundqvist}, {Debnath}, {Delbroek}, {Josiek}, {Lefever}, {Moens}, {Van der Sijpt}, \& {Verhamme}}]{Gonzalez-Tora:25}
{Gonz{\'a}lez-Tor{\`a}}, G., {Sander}, A.~A.~C., {Sundqvist}, J.~O., {et~al.} 2025, \aap, 694, A269, \dodoi{10.1051/0004-6361/202452241}

\bibitem[{{Greco} \& {Danieli}(2022)}]{Greco:22}
{Greco}, J.~P., \& {Danieli}, S. 2022, \apj, 941, 26, \dodoi{10.3847/1538-4357/ac75b7}

\bibitem[{Harris \& Zaritsky(2009)}]{Harris:09}
Harris, J., \& Zaritsky, D. 2009, The Astronomical Journal, 138, 1243, \dodoi{10.1088/0004-6256/138/5/1243}

\bibitem[{{Heckman} {et~al.}(1990){Heckman}, {Armus}, \& {Miley}}]{Heckman:1990}
{Heckman}, T.~M., {Armus}, L., \& {Miley}, G.~K. 1990, \apjs, 74, 833, \dodoi{10.1086/191522}

\bibitem[{{Humphreys} \& {Davidson}(1984)}]{Humphreys:1984}
{Humphreys}, R.~M., \& {Davidson}, K. 1984, Science, 223, 243, \dodoi{10.1126/science.223.4633.243}

\bibitem[{{Iglesias} \& {Rogers}(1993)}]{Iglesias1993}
{Iglesias}, C.~A., \& {Rogers}, F.~J. 1993, \apj, 412, 752, \dodoi{10.1086/172958}

\bibitem[{{Jacobson-Gal{\'a}n} {et~al.}(2024){Jacobson-Gal{\'a}n}, {Dessart}, {Davis}, {Kilpatrick}, {Margutti}, {Foley}, {Chornock}, {Terreran}, {Hiramatsu}, {Newsome}, {Padilla Gonzalez}, {Pellegrino}, {Howell}, {Filippenko}, {Anderson}, {Angus}, {Auchettl}, {Bostroem}, {Brink}, {Cartier}, {Coulter}, {de Boer}, {Drout}, {Earl}, {Ertini}, {Farah}, {Farias}, {Gall}, {Gao}, {Gerlach}, {Guo}, {Haynie}, {Hosseinzadeh}, {Ibik}, {Jha}, {Jones}, {Langeroodi}, {LeBaron}, {Magnier}, {Piro}, {Raimundo}, {Rest}, {Rest}, {Rich}, {Rojas-Bravo}, {Sears}, {Taggart}, {Villar}, {Wainscoat}, {Wang}, {Wasserman}, {Yan}, {Yang}, {Zhang}, \& {Zheng}}]{JacobsonGalan2024}
{Jacobson-Gal{\'a}n}, W.~V., {Dessart}, L., {Davis}, K.~W., {et~al.} 2024, arXiv e-prints, arXiv:2403.02382, \dodoi{10.48550/arXiv.2403.02382}

\bibitem[{{Jermyn} {et~al.}(2023){Jermyn}, {Bauer}, {Schwab}, {Farmer}, {Ball}, {Bellinger}, {Dotter}, {Joyce}, {Marchant}, {Mombarg}, {Wolf}, {Sunny Wong}, {Cinquegrana}, {Farrell}, {Smolec}, {Thoul}, {Cantiello}, {Herwig}, {Toloza}, {Bildsten}, {Townsend}, \& {Timmes}}]{Jermyn:23}
{Jermyn}, A.~S., {Bauer}, E.~B., {Schwab}, J., {et~al.} 2023, \apjs, 265, 15, \dodoi{10.3847/1538-4365/acae8d}

\bibitem[{{Jiang} {et~al.}(2015){Jiang}, {Cantiello}, {Bildsten}, {Quataert}, \& {Blaes}}]{Jiang:15}
{Jiang}, Y.-F., {Cantiello}, M., {Bildsten}, L., {Quataert}, E., \& {Blaes}, O. 2015, \apj, 813, 74, \dodoi{10.1088/0004-637X/813/1/74}

\bibitem[{{Jiang} {et~al.}(2018){Jiang}, {Cantiello}, {Bildsten}, {Quataert}, {Blaes}, \& {Stone}}]{Jiang:18}
{Jiang}, Y.-F., {Cantiello}, M., {Bildsten}, L., {et~al.} 2018, \nat, 561, 498, \dodoi{10.1038/s41586-018-0525-0}

\bibitem[{{Johnston} {et~al.}(2024){Johnston}, {Michielsen}, {Anders}, {Renzo}, {Cantiello}, {Marchant}, {Goldberg}, {Townsend}, {Sabhahit}, \& {Jermyn}}]{Johnston:24}
{Johnston}, C., {Michielsen}, M., {Anders}, E.~H., {et~al.} 2024, \apj, 964, 170, \dodoi{10.3847/1538-4357/ad2343}

\bibitem[{{Kroupa}(2001)}]{Kroupa:01}
{Kroupa}, P. 2001, \mnras, 322, 231, \dodoi{10.1046/j.1365-8711.2001.04022.x}

\bibitem[{{Lan{\c{c}}on} {et~al.}(2007){Lan{\c{c}}on}, {Gallagher}, {de Grijs}, {Hauschildt}, {Ladjal}, {Mouhcine}, {Smith}, {Wood}, \& {Schreiber}}]{Lancon:07}
{Lan{\c{c}}on}, A., {Gallagher}, J.~S., {de Grijs}, R., {et~al.} 2007, in IAU Symposium, Vol. 241, Stellar Populations as Building Blocks of Galaxies, ed. A.~{Vazdekis} \& R.~{Peletier}, 152--155, \dodoi{10.1017/S1743921307007673}

\bibitem[{{Ma} {et~al.}(2025){Ma}, {Justham}, {Pakmor}, {Chiavassa}, {Ryu}, \& {de Mink}}]{Ma:2025}
{Ma}, J.-Z., {Justham}, S., {Pakmor}, R., {et~al.} 2025, arXiv e-prints, arXiv:2510.14875, \dodoi{10.48550/arXiv.2510.14875}

\bibitem[{{Mapelli}(2016)}]{Mapelli:16}
{Mapelli}, M. 2016, \memsai, 87, 567, \dodoi{10.48550/arXiv.1606.03370}

\bibitem[{{Massey} {et~al.}(2021){Massey}, {Neugent}, {Levesque}, {Drout}, \& {Courteau}}]{Massey:21}
{Massey}, P., {Neugent}, K.~F., {Levesque}, E.~M., {Drout}, M.~R., \& {Courteau}, S. 2021, \aj, 161, 79, \dodoi{10.3847/1538-3881/abd01f}

\bibitem[{{Massey} \& {Olsen}(2003)}]{Massey:03}
{Massey}, P., \& {Olsen}, K.~A.~G. 2003, \aj, 126, 2867, \dodoi{10.1086/379558}

\bibitem[{{Mauron} \& {Josselin}(2011)}]{Mauron:11}
{Mauron}, N., \& {Josselin}, E. 2011, \aap, 526, A156, \dodoi{10.1051/0004-6361/201013993}

\bibitem[{{Meynet} \& {Maeder}(2003)}]{Meynet:2003}
{Meynet}, G., \& {Maeder}, A. 2003, \aap, 404, 975, \dodoi{10.1051/0004-6361:20030512}

\bibitem[{{Meynet} {et~al.}(2015){Meynet}, {Chomienne}, {Ekstr{\"o}m}, {Georgy}, {Granada}, {Groh}, {Maeder}, {Eggenberger}, {Levesque}, \& {Massey}}]{Meynet:15}
{Meynet}, G., {Chomienne}, V., {Ekstr{\"o}m}, S., {et~al.} 2015, \aap, 575, A60, \dodoi{10.1051/0004-6361/201424671}

\bibitem[{{Moriya} \& {Yoon}(2022)}]{Moriya:2022}
{Moriya}, T.~J., \& {Yoon}, S.-C. 2022, \mnras, 513, 5606, \dodoi{10.1093/mnras/stac1271}

\bibitem[{{Morozova} {et~al.}(2017){Morozova}, {Piro}, \& {Valenti}}]{Morozova2017}
{Morozova}, V., {Piro}, A.~L., \& {Valenti}, S. 2017, \apj, 838, 28, \dodoi{10.3847/1538-4357/aa6251}

\bibitem[{{Neugent} {et~al.}(2020){Neugent}, {Massey}, {Georgy}, {Drout}, {Mommert}, {Levesque}, {Meynet}, \& {Ekstr{\"o}m}}]{Neugent:20}
{Neugent}, K.~F., {Massey}, P., {Georgy}, C., {et~al.} 2020, \apj, 889, 44, \dodoi{10.3847/1538-4357/ab5ba0}

\bibitem[{{Neugent} {et~al.}(2012){Neugent}, {Massey}, {Skiff}, \& {Meynet}}]{Neugent:12}
{Neugent}, K.~F., {Massey}, P., {Skiff}, B., \& {Meynet}, G. 2012, \apj, 749, 177, \dodoi{10.1088/0004-637X/749/2/177}

\bibitem[{{Nieuwenhuijzen} \& {de Jager}(1990)}]{Nieuwenhuijzen:90}
{Nieuwenhuijzen}, H., \& {de Jager}, C. 1990, \aap, 231, 134

\bibitem[{{Nugis} \& {Lamers}(2000)}]{Nugis:2000}
{Nugis}, T., \& {Lamers}, H.~J.~G.~L.~M. 2000, \aap, 360, 227

\bibitem[{{Owocki} {et~al.}(2004){Owocki}, {Gayley}, \& {Shaviv}}]{Owocki:2004}
{Owocki}, S.~P., {Gayley}, K.~G., \& {Shaviv}, N.~J. 2004, \apj, 616, 525, \dodoi{10.1086/424910}

\bibitem[{{Pastorello} {et~al.}(2007){Pastorello}, {Smartt}, {Mattila}, {Eldridge}, {Young}, {Itagaki}, {Yamaoka}, {Navasardyan}, {Valenti}, {Patat}, {Agnoletto}, {Augusteijn}, {Benetti}, {Cappellaro}, {Boles}, {Bonnet-Bidaud}, {Botticella}, {Bufano}, {Cao}, {Deng}, {Dennefeld}, {Elias-Rosa}, {Harutyunyan}, {Keenan}, {Iijima}, {Lorenzi}, {Mazzali}, {Meng}, {Nakano}, {Nielsen}, {Smoker}, {Stanishev}, {Turatto}, {Xu}, \& {Zampieri}}]{Pastorello:2007}
{Pastorello}, A., {Smartt}, S.~J., {Mattila}, S., {et~al.} 2007, \nat, 447, 829, \dodoi{10.1038/nature05825}

\bibitem[{{Paxton} {et~al.}(2011){Paxton}, {Bildsten}, {Dotter}, {Herwig}, {Lesaffre}, \& {Timmes}}]{Paxton:11}
{Paxton}, B., {Bildsten}, L., {Dotter}, A., {et~al.} 2011, \apjs, 192, 3, \dodoi{10.1088/0067-0049/192/1/3}

\bibitem[{{Paxton} {et~al.}(2013){Paxton}, {Cantiello}, {Arras}, {Bildsten}, {Brown}, {Dotter}, {Mankovich}, {Montgomery}, {Stello}, {Timmes}, \& {Townsend}}]{paxton:13}
{Paxton}, B., {Cantiello}, M., {Arras}, P., {et~al.} 2013, \apjs, 208, 4, \dodoi{10.1088/0067-0049/208/1/4}

\bibitem[{{Paxton} {et~al.}(2015){Paxton}, {Marchant}, {Schwab}, {Bauer}, {Bildsten}, {Cantiello}, {Dessart}, {Farmer}, {Hu}, {Langer}, {Townsend}, {Townsley}, \& {Timmes}}]{paxton:15}
{Paxton}, B., {Marchant}, P., {Schwab}, J., {et~al.} 2015, \apjs, 220, 15, \dodoi{10.1088/0067-0049/220/1/15}

\bibitem[{{Paxton} {et~al.}(2018){Paxton}, {Schwab}, {Bauer}, {Bildsten}, {Blinnikov}, {Duffell}, {Farmer}, {Goldberg}, {Marchant}, {Sorokina}, {Thoul}, {Townsend}, \& {Timmes}}]{paxton:18}
{Paxton}, B., {Schwab}, J., {Bauer}, E.~B., {et~al.} 2018, \apjs, 234, 34, \dodoi{10.3847/1538-4365/aaa5a8}

\bibitem[{{Paxton} {et~al.}(2019){Paxton}, {Smolec}, {Schwab}, {Gautschy}, {Bildsten}, {Cantiello}, {Dotter}, {Farmer}, {Goldberg}, {Jermyn}, {Kanbur}, {Marchant}, {Thoul}, {Townsend}, {Wolf}, {Zhang}, \& {Timmes}}]{paxton:19}
{Paxton}, B., {Smolec}, R., {Schwab}, J., {et~al.} 2019, \apjs, 243, 10, \dodoi{10.3847/1538-4365/ab2241}

\bibitem[{{Pereira} {et~al.}(2024){Pereira}, {Janot-Pacheco}, {Emilio}, {Andrade}, {Armstrong}, {Eidam}, {Rabello-Soares}, \& {da Silva}}]{Pereira:24}
{Pereira}, A.~W., {Janot-Pacheco}, E., {Emilio}, M., {et~al.} 2024, \aap, 686, A20, \dodoi{10.1051/0004-6361/202346439}

\bibitem[{Rahmani {et~al.}(2016)Rahmani, Lianou, \& Barmby}]{Rahmani:16}
Rahmani, S., Lianou, S., \& Barmby, P. 2016, Monthly Notices of the Royal Astronomical Society, 456, 4128, \dodoi{10.1093/mnras/stv2951}

\bibitem[{{Ren} {et~al.}(2021){Ren}, {Jiang}, {Yang}, {Wang}, \& {Ren}}]{Ren:21}
{Ren}, Y., {Jiang}, B., {Yang}, M., {Wang}, T., \& {Ren}, T. 2021, \apj, 923, 232, \dodoi{10.3847/1538-4357/ac307b}

\bibitem[{{Renzo} {et~al.}(2017){Renzo}, {Ott}, {Shore}, \& {de Mink}}]{Renzo:17}
{Renzo}, M., {Ott}, C.~D., {Shore}, S.~N., \& {de Mink}, S.~E. 2017, \aap, 603, A118, \dodoi{10.1051/0004-6361/201730698}

\bibitem[{{Russell} \& {Dopita}(1992)}]{Russell:92}
{Russell}, S.~C., \& {Dopita}, M.~A. 1992, \apj, 384, 508, \dodoi{10.1086/170893}

\bibitem[{{Sanders} {et~al.}(2012){Sanders}, {Caldwell}, {McDowell}, \& {Harding}}]{Sanders:12}
{Sanders}, N.~E., {Caldwell}, N., {McDowell}, J., \& {Harding}, P. 2012, \apj, 758, 133, \dodoi{10.1088/0004-637X/758/2/133}

\bibitem[{{Schlegel}(1990)}]{Schlegel:1990}
{Schlegel}, E.~M. 1990, \mnras, 244, 269

\bibitem[{{Shiber} {et~al.}(2024){Shiber}, {Chatzopoulos}, {Munson}, \& {Frank}}]{Shiber:24}
{Shiber}, S., {Chatzopoulos}, E., {Munson}, B., \& {Frank}, J. 2024, \apj, 962, 168, \dodoi{10.3847/1538-4357/ad0e0a}

\bibitem[{{Smith}(2014)}]{Smith:2014}
{Smith}, N. 2014, \araa, 52, 487, \dodoi{10.1146/annurev-astro-081913-040025}

\bibitem[{{Smith}(2017)}]{Smith:2017}
---. 2017, Philosophical Transactions of the Royal Society of London Series A, 375, 20160268, \dodoi{10.1098/rsta.2016.0268}

\bibitem[{{Smith} \& {Owocki}(2006)}]{Smith:2006}
{Smith}, N., \& {Owocki}, S.~P. 2006, \apjl, 645, L45, \dodoi{10.1086/506523}

\bibitem[{{Smith} {et~al.}(2004){Smith}, {Vink}, \& {de Koter}}]{Smith:04}
{Smith}, N., {Vink}, J.~S., \& {de Koter}, A. 2004, \apj, 615, 475, \dodoi{10.1086/424030}

\bibitem[{{Steinwandel} \& {Goldberg}(2023)}]{Steinwandel:2023}
{Steinwandel}, U.~P., \& {Goldberg}, J.~A. 2023, arXiv e-prints, arXiv:2310.11495, \dodoi{10.48550/arXiv.2310.11495}

\bibitem[{{Tacconi} {et~al.}(2020){Tacconi}, {Genzel}, \& {Sternberg}}]{Tacconi:20}
{Tacconi}, L.~J., {Genzel}, R., \& {Sternberg}, A. 2020, \araa, 58, 157, \dodoi{10.1146/annurev-astro-082812-141034}

\bibitem[{{Uchida} {et~al.}(2019){Uchida}, {Shibata}, {Takahashi}, \& {Yoshida}}]{Uchida:2019}
{Uchida}, H., {Shibata}, M., {Takahashi}, K., \& {Yoshida}, T. 2019, \prd, 99, 041302, \dodoi{10.1103/PhysRevD.99.041302}

\bibitem[{{van Loon} {et~al.}(2005){van Loon}, {Cioni}, {Zijlstra}, \& {Loup}}]{vanLoon:05}
{van Loon}, J.~T., {Cioni}, M. R.~L., {Zijlstra}, A.~A., \& {Loup}, C. 2005, \aap, 438, 273, \dodoi{10.1051/0004-6361:20042555}

\bibitem[{{Vink} \& {de Koter}(2002)}]{Vink:02}
{Vink}, J.~S., \& {de Koter}, A. 2002, \aap, 393, 543, \dodoi{10.1051/0004-6361:20021009}

\bibitem[{{Vink} {et~al.}(2001){Vink}, {de Koter}, \& {Lamers}}]{Vink:2001}
{Vink}, J.~S., {de Koter}, A., \& {Lamers}, H.~J.~G.~L.~M. 2001, \aap, 369, 574, \dodoi{10.1051/0004-6361:20010127}

\bibitem[{{Vink} {et~al.}(2021){Vink}, {Higgins}, {Sander}, \& {Sabhahit}}]{Vink:21}
{Vink}, J.~S., {Higgins}, E.~R., {Sander}, A. A.~C., \& {Sabhahit}, G.~N. 2021, \mnras, 504, 146, \dodoi{10.1093/mnras/stab842}

\bibitem[{{Vink} \& {Sabhahit}(2023)}]{Vink:23}
{Vink}, J.~S., \& {Sabhahit}, G.~N. 2023, \aap, 678, L3, \dodoi{10.1051/0004-6361/202347801}

\bibitem[{{Wood} {et~al.}(1983){Wood}, {Bessell}, \& {Fox}}]{Wood:83}
{Wood}, P.~R., {Bessell}, M.~S., \& {Fox}, M.~W. 1983, \apj, 272, 99, \dodoi{10.1086/161265}

\bibitem[{{Wood} {et~al.}(1992){Wood}, {Whiteoak}, {Hughes}, {Bessell}, {Gardner}, \& {Hyland}}]{Wood:92}
{Wood}, P.~R., {Whiteoak}, J.~B., {Hughes}, S.~M.~G., {et~al.} 1992, \apj, 397, 552, \dodoi{10.1086/171812}

\bibitem[{{Yang} {et~al.}(2020){Yang}, {Bonanos}, {Jiang}, {Gao}, {Gavras}, {Maravelias}, {Wang}, {Chen}, {Tramper}, {Ren}, {Spetsieri}, \& {Xue}}]{Yang:20}
{Yang}, M., {Bonanos}, A.~Z., {Jiang}, B.-W., {et~al.} 2020, \aap, 639, A116, \dodoi{10.1051/0004-6361/201937168}

\bibitem[{{Yang} {et~al.}(2023){Yang}, {Bonanos}, {Jiang}, {Zapartas}, {Gao}, {Ren}, {Lam}, {Wang}, {Maravelias}, {Gavras}, {Wang}, {Chen}, {Tramper}, {de Wit}, {Chen}, {Wen}, {Liu}, {Tian}, {Antoniadis}, \& {Luo}}]{Yang:23}
{Yang}, M., {Bonanos}, A.~Z., {Jiang}, B., {et~al.} 2023, \aap, 676, A84, \dodoi{10.1051/0004-6361/202244770}

\bibitem[{Yang {et~al.}(2023)Yang, Bonanos, Jiang, Zapartas, Gao, Ren, Lam, Wang, Maravelias, Gavras, Wang, Chen, Tramper, De~Wit, Chen, Wen, Liu, Tian, Antoniadis, \& Luo}]{10.26093/cds/vizier.36760084}
Yang, M., Bonanos, A., Jiang, B., {et~al.} 2023, Mass-loss rate of RSGs in the SMC,  Centre de Donnees Strasbourg (CDS), \dodoi{10.26093/CDS/VIZIER.36760084}

\bibitem[{{Zapartas} {et~al.}(2025){Zapartas}, {de Wit}, {Antoniadis}, {Mu{\~n}oz-Sanchez}, {Souropanis}, {Bonanos}, {Maravelias}, {Kovlakas}, {Kruckow}, {Fragos}, {Andrews}, {Bavera}, {Briel}, {Gossage}, {Kasdagli}, {Rocha}, {Sun}, {Srivastava}, \& {Xing}}]{Zapartas:25}
{Zapartas}, E., {de Wit}, S., {Antoniadis}, K., {et~al.} 2025, \aap, 697, A167, \dodoi{10.1051/0004-6361/202452401}

\bibitem[{{Ziosi} {et~al.}(2014){Ziosi}, {Mapelli}, {Branchesi}, \& {Tormen}}]{Ziosi:14}
{Ziosi}, B.~M., {Mapelli}, M., {Branchesi}, M., \& {Tormen}, G. 2014, \mnras, 441, 3703, \dodoi{10.1093/mnras/stu824}

\end{thebibliography}
\bibliographystyle{aasjournal}
\end{document}